\documentclass[a4paper,11pt]{article}
\pdfoutput=1 

\usepackage{jheppub} 

\usepackage{amsmath, mathtools}
\usepackage{amsfonts}
\usepackage{amssymb}
\usepackage{accents}
\usepackage{color}
\usepackage{slashed}
\usepackage{comment}
\usepackage{hyperref}
\usepackage{booktabs}
\usepackage{braket}
\usepackage{cellspace}
\usepackage{tikz}
\usepackage{pgfplots}
\usepackage{xcolor}

\usepgfplotslibrary{external}

\numberwithin{equation}{section}

\newcommand{\AdS}[1]{\text{AdS}_{#1}}
\newcommand{\CFT}{\text{CFT}}

\newcommand{\de}{\text{d}}

\newcommand{\var}{x}

\newcommand {\cD}{{\cal D}}

\newcommand {\cF}{{\cal F}}

\newcommand {\cH}{{\cal H}}

\newcommand {\cJ}{{\cal J}}
\newcommand {\cK}{{\cal K}}

\newcommand {\cN}{{\cal N}}
\newcommand {\cO}{{\cal O}}

\newcommand {\cQ}{{\cal Q}}

\newcommand {\cS}{{\cal S}}
\newcommand {\cT}{{\cal T}}

\newcommand {\cX}{{\cal X}}

\def\a{\alpha}

\def\d{\delta}
\def\e{\epsilon}

\def\s{\sigma}

\def\z{\zeta}

\def\F{\Phi}

\def\ri{{\rm i}}

\newcommand{\pa}{\partial}
\newcommand{\hf}{\frac12}

\newcommand{\be}{\begin{equation}}
\newcommand{\ee}{\end{equation}}
\newcommand{\bea}{\begin{eqnarray}}
\newcommand{\eea}{\end{eqnarray}}

\newcommand{\ba}{\begin{array}}
\newcommand{\ea}{\end{array}}

\def\double #1{#1{\hbox{\kern-2pt $#1$}}}

\newcommand{\bsubeq}{\begin{subequations}}
\newcommand{\esubeq}{\end{subequations}}

\newcommand{\rd}{\text{d}}
\numberwithin{equation}{section}  

\title{ On \boldmath $T\bar{T}$ deformations and supersymmetry}

\author[a]{Marco Baggio,}
\author[b]{Alessandro Sfondrini,}
\author[a,c]{Gabriele Tartaglino-Mazzucchelli,}
\author[b]{Harriet Walsh}

\affiliation[a]{Instituut voor Theoretische Fysica, KU Leuven,\\
Celestijnenlaan 200D, B-3001 Leuven, Belgium}
\affiliation[b]{Institut f\"ur theoretische Physik, ETH Z\"urich\\ Wolfgang-Pauli-Stra{\ss}e 27, 8093 Z\"urich, Switzerland}
\affiliation[c]{Albert Einstein Center for Fundamental Physics,
Institute for Theoretical Physics,\\
University of Bern,
Sidlerstrasse 5, CH-3012 Bern, Switzerland}

\emailAdd{marco.baggio@kuleuven.be}
\emailAdd{sfondria@itp.phys.ethz.ch}
\emailAdd{gtm@itp.unibe.ch}
\emailAdd{walshh@student.ethz.ch}

\abstract{We investigate the ``$T\bar{T}$'' deformations of two-dimensional supersymmetric quantum field theories. 
More precisely, we show that, by using the conservation equations for the supercurrent multiplet,
the $T\bar{T}$ deforming operator can be constructed as a supersymmetric descendant.
Here we focus 
on $\mathcal{N}=(1,0)$
and 
$\mathcal{N}=(1,1)$ supersymmetry.
As an example,
we analyse in detail the $T\bar{T}$ deformation of a free $\mathcal{N}=(1,0)$ supersymmetric action.
We also argue that the link between $T\bar{T}$ and string theory can be extended to superstrings:
by analysing the light-cone gauge fixing for superstrings in flat space, we show the correspondence of the string 
action to the $T\bar{T}$ deformation of  a free theory of eight  $\mathcal{N}=(1,1)$ scalar multiplets on the nose. 
We comment on how these constructions relate to the geometrical interpretations of $T\bar{T}$ deformations that 
have recently been discussed in the literature.
}

\begin{document}
\maketitle
\flushbottom

\newpage

\section{Introduction}
The deformations arising from the composite ``$T\bar{T}$'' operator~\cite{Zamolodchikov:2004ce}, constructed out of 
the (Hilbert) stress-energy tensor, have recently attracted much attention. These irrelevant deformations are solvable, 
in the sense that they affect the theory's energy spectrum in a simple way.
For a state $|n\rangle$ of energy $H_n$ 
and zero momentum, a $T\bar{T}$ deformation of parameter~$\alpha$ acts as
\begin{equation}
\label{eq:TTbargeneral}
\partial_\alpha H_n= -\langle n | O|n\rangle = 
H_n\partial_RH_n\,,
\end{equation}
where $O=\text{det}[T^{ab}]$ is the $T\bar{T}$ composite operator and~$R$ is the volume of the theory. This means 
that the energy of a state $|n\rangle$ in the $\alpha$-deformed theory in volume~$R_0$ is the same as that of the 
same state in the undeformed theory, but in volume $R=R_0+\alpha H_n$.
For integrable theories, such a deformation is identified with a ``CDD factor''~\cite{Castillejo:1955ed} in the factorised 
S~matrix~\cite{Smirnov:2016lqw,Cavaglia:2016oda}. 
This CDD factor appears to be closely related to (effective) string theories on flat space~\cite{Dubovsky:2012wk,Dubovsky:2012sh, Caselle:2013dra,Cavaglia:2016oda,Chen:2018keo} and stringy Wess-Zumino-Witten models~\cite{Baggio:2018gct,Dei:2018mfl,Dei:2018jyj}, and more generally with string theories in uniform light-cone gauge~\cite{Arutyunov:2004yx,Arutyunov:2005hd, Arutyunov:2006gs,Baggio:2018gct}.%
\footnote{See also ref.~\cite{Arutyunov:2009ga} for a review of (integrable) strings in uniform light-cone gauge.}
 The holographic interpretation of such deformations has also been directly investigated~\cite{McGough:2016lol,Giveon:2017nie, Giveon:2017myj, Giribet:2017imm, Kraus:2018xrn}, as well as their relation with two-dimensional (Jackiw-Teitelboim~\cite{Jackiw:1984je,Teitelboim:1983ux}) gravitational theories~\cite{Dubovsky:2017cnj,Cardy:2018sdv, Dubovsky:2018bmo,Conti:2018tca}.

One aspect that has been perhaps overlooked so far is how $T\bar{T}$ deformations affect a supersymmetric theory. This is a rather natural question given that two-dimensional supersymmetric theories are of great interest, both in their own right and in the context of string theory and holography. In this short note we aim to address this question. 

If we think of the deformation as arising from a CDD factor that yields the differential equation~\eqref{eq:TTbargeneral}, it is natural to conclude that it should behave well with respect to supersymmetry. States related to each other by the action of supersymmetry will have the same energy and momentum. As long as the flow of eq.~\eqref{eq:TTbargeneral} is non-singular, such degeneracies should be preserved. This seems to imply that a supersymmetric theory would generically remain so after a $T\bar{T}$ deformation.

However, to have complete control over the deformed theory---rather than on its spectrum only---it is important to construct the $T\bar{T}$ operator, which we will simply denote in the paper as
$O=\text{det}[T^{ab}]$, and determine how it behaves with respect to supersymmetry. It is easy to see that, even for a very simple (free) theory, $\text{det}[T^{ab}]$ is not invariant under the supersymmetry variations of the undeformed theory.
In this paper we will argue that it is possible to 
define a deformed $T\bar{T}$ operator that \textit{is supersymmetric} with respect to the supersymmetries of the undeformed theory and \textit{is equivalent to} $O$, up to total derivatives and terms that vanish when using the conservation equations of the supercurrents  (which hold on the mass shell). 

More specifically, we will consider $\mathcal{N}=(1,0)$ and $\mathcal{N}=(1,1)$ theories,
and show that $O$ can be constructed as a supersymmetric descendant
of an appropriate composite operator. For instance in the $\mathcal{N}=(1,0)$ case we have
\begin{equation}
\label{eq:TTbarIsDescendant}
\int \de^2\sigma\, O(\s) = \int \de^2\sigma \big\{{Q}_+, {O}_{-}(\s)\big\}\,, 
\end{equation}
\textit{up to terms that vanish on-shell},
where ${Q}_+$ is the supercharge and ${O}_{-}(\s)$ a suitable fermionic operator. This is sufficient to guarantee that
\begin{equation}
\Big[{Q}_+\,,\,\int \de^2\sigma\, O(\s)\Big]=0\,,
\end{equation}
so that the deformed action remains supersymmetric.
The fact that eq.~\eqref{eq:TTbarIsDescendant} is corrected by terms that vanish on-shell is inconsequential. Indeed we will argue that such terms do not affect expectation values, so that the two operators on either side of eq.~\eqref{eq:TTbarIsDescendant} can be defined by point splitting, and will generate the same exactly solvable flow as eq.~\eqref{eq:TTbargeneral}.

We derive eq.~\eqref{eq:TTbarIsDescendant} and discuss these ideas
for $\cN=(1,0)$ and $\cN=(1,1)$ supersymmetry in section~\ref{sec:TTbarsusy}.
In section~\ref{sec:TTbarN10} we provide a concrete example by discussing in detail the deformation of the free 
$\mathcal{N}=(1,0)$ supersymmetric action for a scalar multiplet.
A particularly interesting feature of $T\bar{T}$ deformations is that they appear to be naturally related to string theory. This was first observed for bosonic strings in flat space~\cite{Dubovsky:2012wk,Dubovsky:2012sh,Caselle:2013dra, Cavaglia:2016oda,Chen:2018keo}.
In ref.~\cite{Baggio:2018gct} it was argued that such a relation is  more general, and should affect any string theory quantised in uniform light-cone gauge~\cite{Arutyunov:2004yx,Arutyunov:2005hd,Arutyunov:2006gs}, see also~\cite{Arutyunov:2009ga}. Indeed, in ref.~\cite{Baggio:2018gct} such relation was exploited to study the integrable structure of AdS$_3$ string backgrounds with Neveu-Schwarz-Neveu-Schwarz background fluxes.%
\footnote{%
See also ref.~\cite{Sfondrini:2014via} for a review of integrability of AdS$_3$ superstring backgrounds.}
In section~\ref{sec:TTbarstring} we argue that this relation holds also for \textit{supersymmetric} string theories, and we show explicitly that this is the case for Green-Schwarz superstrings in flat space. In addition, we comment on how our construction, which is based on ``uniform'' light-cone gauge~\cite{Arutyunov:2004yx,Arutyunov:2005hd,Arutyunov:2006gs}, relates to other approaches to the~$T\bar{T}$ deformation.
We conclude with a brief discussion in section~\ref{sec:conclusions}, and collect a number of technical results in the appendices.

\paragraph{Note added.}
The pre-print~\cite{Chang:2018dge}, which appeared on the arXiv shortly after this pre-print, also addresses the relation between $T\bar{T}$ deformations and supersymmetry. The results presented there are compatible with the discussion of our sections~\ref{sec:TTbarsusy} and~\ref{sec:TTbarN10}.


\section{The  \texorpdfstring{$T\bar{T}$}{TTbar}  operator as a supersymmetric descendant}
\label{sec:TTbarsusy}

In this section we will provide some general arguments showing that the $T\bar{T}$ operator 
of theories possessing two-dimensional (2D) supersymmetry is itself supersymmetric. Specifically, we will see that it is the supersymmetric descendant of a composite operator
constructed out of elements of the supercurrent multiplet.

We will restrict our analysis to the case of two-dimensional $\cN=(1,0)$ and $\cN=(1,1)$ supersymmetry.
Moreover, we will assume that the supersymmetric QFTs in consideration 
are Lorentz invariant and can be coupled to two-dimensional off-shell supergravity, and
to simplify our analysis we shall work in superspace.
The study of 2D $(1,0)$ and $(1,1)$ supergravity in superspace was largely developed in the 1980s
and we refer the reader to the following works and references therein for details 
\cite{Gates:1984nk,Gates:1985vk,Brooks:1986hv,Brooks:1986bc,Brooks:1986gd,Brooks:1986uh,Gates:1986ez,Brooks:1986mv,Smailagic:1992pj}.
Under our assumptions, the description of the supercurrent multiplet will simplify: the supercurrent and trace multiplet will coincide with the variational derivatives of the supersymmetric matter action 
with respect to the supergravity prepotentials, evaluated on a Minkowski superspace background
(see \cite{Gates:1983nr,Buchbinder:1998qv} for general reviews on the construction of supercurrents
in superspace along this line). 
This is the supergravity analogue to the definition of the Hilbert stress-energy tensor as the functional derivative of 
a matter theory minimally coupled to a background metric (vielbein).
We leave for future work the extension of our arguments to more general supercurrent multiplets derived
in the spirit of the analysis of, 
\textit{e.g.}\ refs.~\cite{Magro:2001aj,Komargodski:2010rb,Kuzenko:2010am,Dumitrescu:2011iu}.

\subsection{The \texorpdfstring{$\cN=(1,0)$}{N=(1,0)} supercurrent multiplet}

We start by introducing the $\cN=(1,0)$ superspace and the structure of the supercurrent multiplet.
In light-cone coordinates, 
a flat 2D $\cN=(1,0)$ superspace is parametrised by (see appendix \ref{app:conventions2d} for our definition
of $\s^{\pm\pm}$)
\begin{equation}
\z^M=(\s^{++},\s^{--},\vartheta^+)\,,
\end{equation}
with $\vartheta^+$ a real Grassmann coordinate.
The spinor covariant derivatives and supercharges are given by
\bea
\cD_+ =  \frac{\pa}{\pa \vartheta^+} - {\rm i}  \vartheta^+ \pa_{++}
\,,\qquad
\cQ_+ 
= {\rm i}\, \frac{\pa}{\pa \vartheta^+} -   \vartheta^+ \pa_{++}
~,
\eea
and obey the anti-commutation relation
\be
\{ \cD_+ , \cD_+ \} =-2 \ri \pa_{++}\,,\qquad
\{ \cQ_+ , \cQ_+ \} =-2 \ri \pa_{++}\,,\qquad
\{\cQ_+, \cD_+\}=0~.
\ee
Given a $(1,0)$ superfield $\cF(\z)=\cF(\s,\vartheta^+)$ its supersymmetry transformations are given by
\bea
\d_Q \cF:=-\ri \e_- \cQ_+ \cF(\s,\vartheta^+)
~.
\eea
Note also that if $F(\s)$ is the operator defined as the $\vartheta=0$ component of the superfield 
$\cF(\zeta)$, $F(\s):=\cF(\s,\vartheta^+)|_{\vartheta=0}$, then 
its supersymmetry transformations are such that
\bea
\d_Q F(\s)=-\ri\e_-\big[Q_+,F(\s)\big\}=-\ri \e_- \cQ_+ \cF(\s,\vartheta^+)\Big|_{\vartheta=0}
=\e_- \cD_+ \cF(\s,\vartheta^+)\Big|_{\vartheta=0}
~.
\eea
We will indicate by $Q_+$ the supersymmetry generator \textit{acting on a component operator}
and distinguish it from $\cQ_+$, which is a linear operator \textit{acting on superfields}.

The description of general $\cN=(1,0)$ supergravity-matter systems in superspace
has been developed in full detail three decades ago. See \cite{Brooks:1986hv,Brooks:1986bc,Brooks:1986gd,Brooks:1986uh,Gates:1986ez,Brooks:1986mv,Smailagic:1992pj} for an (incomplete) list of references.
We will restrict our attention to a Lorentz invariant Lagrangian matter system 
coupled to linearised off-shell Poincar\'e supergravity.
In this case, linearised supergravity can be described by three unconstrained prepotential superfields:
\begin{equation}
\cH_{+}{}^{--}(\zeta)=-\hf\cH_{+++}(\zeta)\,,\qquad
\cH_{--}{}^{++}(\zeta)=-\hf\cH_{----}(\zeta)\,,\qquad\text{and}\quad \cS(\zeta)\,,
\end{equation}
the last one being a Lorentz scalar.
Their linearised supergravity transformations are
\begin{equation}
\begin{aligned}
\d \cH_{+++}&=\ -\cD_+\cK_{++}
\,,
\\
\d \cH_{----}&=\ \pa_{--}\cK_{--}
\,,
\\
\d \cS&=\ 
\hf(\pa_{++} \cK_{--}+\pa_{--} \cK_{++})+\cK_D
\,,
\end{aligned}
\label{lin-sugra-10}
\end{equation}
where $\cK_{\pm\pm}(\zeta)$ and $\cK_D(\zeta)$  
are unconstrained and parametrise linearised superdiffeomorphisms and
super-Weyl transformations, respectively. 
The supergravity multiplet described here is equivalent to $\mathcal{N}=(1,0)$ conformal supergravity
coupled to a scalar conformal compensator multiplet.
The component fields of the prepotentials are
\begin{equation}
\begin{aligned}
&\cH_{+++}(\zeta)&=&&&\rho_{+++}(\s)&+&&&\ri\vartheta^+\, h_{++++}(\s)
\,,
\\
&\cH_{----}(\zeta)&=&&&h_{----}(\s)&+&&&\ri\vartheta^+\, \psi_{---}(\s)
\,,
&\\
&\cS(\zeta)&=&&&h(\s)&+&&&\ri\vartheta^+\, \psi_{+}(\s)
\,,
\end{aligned}
\end{equation}
where $h_{++++}$ and $h_{----}$ being the traceless components of the linearised metric
while $h$ is its trace, $\psi_{---}$ and $\psi_{+}$ are the gravitini while 
the field $\rho_{+++}$ is pure gauge.

Now consider a general Lorentz invariant matter system coupled to $\mathcal{N}=(1,0)$ supergravity. 
Its action expanded to first order in the supergravity prepotential is
\begin{equation}
S=-\frac{\ri}{8}\int\rd^2\s\,\rd\vartheta^+\Big{[} \cH_{+++}\,\cT_{----}+ \ri \cH_{----}\,\cJ_{+++}+2\ri \cS\,\cJ_-\Big{]}
\,.
\label{action-sources-10}
\end{equation}
Assuming that the equations of motion for the matter are satisfied,
after some integrations by parts
the variation of the action under arbitrary superdiffeomorphism transformations takes the form
\bea
\d S=-\frac{\ri}{8}\int\rd^2\s\,\rd\vartheta^+\Big{[}\cK_{++}\big(\cD_+\cT_{----}-\ri\pa_{--}\cJ_-\big)
-\ri \cK_{--}\big(\pa_{--}\cJ_{+++}+\pa_{++}\cJ_-\big)
\Big{]}
~.~~~
\eea
Imposing that the action is invariant then leads to the following supercurrent conservation equations:
\begin{equation}
\label{supercurrents-10}
\begin{aligned}
\cD_+\cT_{----}&=\ \ri\pa_{--}\cJ_-
\,,
\\
\pa_{--}\cJ_{+++}
&=\ -\pa_{++}\cJ_-
\,.
\end{aligned}
\end{equation}
The on-shell variation of the action \eqref{action-sources-10} under arbitrary super-Weyl 
transformations is
\bea
\d S=\frac{1}{4}\int\rd^2\s\,\rd\vartheta^+\,\cK_D \cJ_-
~.
\label{superconformal-10}
\eea
If the matter system is superconformal, 
which will not be our main interest in our paper,
this variation should vanish
for arbitrary $\cK_D$, which implies that
\bea
\cJ_-(\zeta)=0
\,.
\eea
It is clear that $\cT_{----}$, $\cJ_{+++}$ and $\cJ_-$
belong to a supercurrent multiplet where $\cJ_-$ plays the role of the supertrace, the 
supersymmetric analogue of the 
trace of the stress-energy tensor.

The components of the supercurrent multiplet are
\begin{equation}
\begin{aligned}
&\cJ_{+++}(\zeta)&=&&&J_{+++}(\s)&+&&&\phantom{\ri}\vartheta^+\,T_{++++}(\s)
\,,
\\
&\cT_{----}(\zeta)&=&&&T_{----}(\s)&+&&&\ri\vartheta^+\,\pa_{--}J_{-}(\s)
\,,
\\
&\cJ_-(\zeta)&=&&&J_-(\s)&+&&&\phantom{\ri}\vartheta^+\,\Theta(\s)
\,,
\end{aligned}
\end{equation}
where the fields $T_{++++}(\s)=\cD_+\cJ_{+++}|_{\vartheta=0}$, $T_{----}(\s)=\cT_{----}|_{\vartheta=0}$
and $\Theta(\s)=\cD_+\cJ_-|_{\vartheta=0}$ are the light-cone components of the stress-energy tensor
while $J_{+++}(\s)=\cJ_{+++}|_{\vartheta=0}$ and $J_{-}(\s)=\cJ_{-}|_{\vartheta=0}$ 
are the components of the $\cQ_+$-supersymmetry 
current.
By using \eqref{supercurrents-10},
it is straightforward to show that these operators satisfy the correct conservation equations 
\begin{equation}
\begin{aligned}
&\pa_{++}T_{----}&=&&&-\pa_{--}\Theta\,,\\
&\pa_{--}T_{++++}&=&&&-\pa_{++}\Theta\,,\\
&\pa_{--}J_{+++}&=&&&-\pa_{++}J_-\,.
\end{aligned}
\end{equation}

\subsection{The \texorpdfstring{$T\bar{T}$}{TTbar} operator}

Now that we have described the structure of the supercurrent multiplet, we are ready to 
show that the $T\bar{T}$ operator, which takes the form
\bea
O(\s)=T_{++++}(\s)\,T_{----}(\s)-\big[\Theta(\s)\big]^2
~,
\eea
is a supersymmetric descendant.
Defining the superfields $\cT$ and $\cT_{++++}$ as
\begin{equation}
\cT(\z)\equiv \cD_+\cJ_-(\z),\qquad
\cT_{++++}(\z)\equiv\cD_+\cJ_{+++}(\z)\,
\end{equation}
and using eq.~\eqref{supercurrents-10},
the superfield
\bea
\cO_{-}(\z):=\cT_{----}(\z)\,\cJ_{+++}(\z)-\cT(\z)\,\cJ_-(\z)
~,
\label{susy-TTbar-10}
\eea
is such that
\begin{equation}
\begin{aligned}
\cD_+\cO_{-}(\z)
=\ &
\cT_{++++}(\z)\cT_{----}(\z)
-\big[\cT(\z)\big]^2
\\
&
-\cJ_{+++}(\z)\big[\cD_+\cT_{----}(\z)-\ri\pa_{--}\cJ_{-}(\z)\big]
\\
&
-\ri\cJ_{-}(\z)\big[\pa_{--}\cJ_{+++}(\z)+\pa_{++}\cJ_-(\z)\big]
\\
&
-\ri\pa_{--}\big[\cJ_{+++}(\z)\cJ_{-}(\z)\big]
+\ri\pa_{++}\big[\cJ_-(\z)\cJ_-(\z)\big]
~.
\end{aligned}
\end{equation}
Using the conservation equations \eqref{supercurrents-10}, 
the previous expression becomes
\begin{equation}
\begin{aligned}
\cD_+\cO_{-}(\z)
=\ &
\cT_{++++}(\z)\,\cT_{----}(\z)-\big[\cT(\z)\big]^2\\
&
+\ri\pa_{++}\big[\cJ_-(\z)\,\cJ_-(\z)\big]
-\ri\pa_{--}\big[\cJ_{+++}(\z)\,\cJ_-(\z)\big]\,.
\end{aligned}
\end{equation}
This implies that, up to total derivatives, the following equality holds
\bea
O(\s)
=\int\rd\vartheta^+\,\cO_{-}(\z)
=\cD_+\cO_{-}(\z)\Big|_{\vartheta=0}
=
\ri\big\{Q_+,O_{-}(\s)\big\}
~,
\eea
where we have defined
\begin{equation}
O_{-}(\s)\equiv \cO_{-}(\z)\Big|_{\vartheta=0}\,.
\end{equation}
In other words, $O_-(\s)$ is the supersymmetric primary field of the multiplet containing the operator~$O(\s)$.
As we anticipated in the introduction, this shows that the $T\bar{T}$ deformation is manifestly supersymmetric, since
\bea
{\Big[}Q_+\,,\,\int\rd^2\s\, O(\s){\Big]}
=
\ri\int\rd^2\s{\Big[}Q_+\,,\, \big\{Q_+\,,\,O_-(\s)\big\}{\Big]}
= 0
~.
\eea

It is also rather interesting to write down the deformed action, with deformation parameter~$\alpha$, in terms of the original action $S_0\equiv S$ of eq.~\eqref{action-sources-10} and of the operator $O(\sigma)$ defined above:
\begin{equation}
\begin{aligned}
S_{\alpha}=\ & S_0-\alpha\,\int \de^2\sigma \, O(\sigma)
+\cdots\\
=\ &-\frac{\ri}{8} \int\de^2\sigma\de\vartheta^+\Big[
\big(\mathcal{H}_{+++}-4\ri\alpha\mathcal{J}_{+++}\big)\mathcal{T}_{----}
\\
&~~~~~~~~~~~~~~~~~~~~~
+\ri\big(\mathcal{H}_{----}-4\alpha\mathcal{T}_{----}\big)\mathcal{J}_{+++}
\\
&~~~~~~~~~~~~~~~~~~~~~
+2\mathrm{i}\big(\mathcal{S}+4\alpha\mathcal{T}\big)\,\mathcal{J}_{-}
\Big]
+\cdots\,,
\end{aligned}
\end{equation}
where the ellipses refer to higher order terms in $\a$.
In other words, the $T\bar{T}$ deformation can be reabsorbed in a shift of the prepotential superfields; this will be the case also for $\mathcal{N}=(1,1)$ theories, as we will see below. This points towards a geometric interpretation of such a deformation in terms of the supergravity, reminiscent of the observations of refs.~\cite{Dubovsky:2017cnj,Dubovsky:2018bmo,Conti:2018tca}, which would be interesting to explore further.

\subsection{Point splitting}
\label{sec:onshell}
The composite operator defined in eq.~\eqref{susy-TTbar-10} is well-defined via point splitting, as proven in ref.~\cite{Smirnov:2016lqw}. Indeed, consider the point-split superfield operator
\bea
\cO_{-}(\zeta,\zeta'):=\cT_{----}(\zeta)\,\cJ_{+++}(\zeta')-\cT(\zeta)\,\cJ_-(\zeta')
~.
\label{composite-superfield}
\eea
By expanding it in powers of the fermionic coordinates $\vartheta$ and $\vartheta'$, it is straightforward to check that each component is of the form
\bea
A_{s}(\sigma)\, A'_{s'}(\sigma')-B_{s+2}(\sigma)\, B_{s'-2}(\sigma')~,
\label{AABB}
\eea
where $s$ and $s'$ label the spins of the operators and
\bea
\partial_{++} A_{s} &= - \partial_{--} B_{s+2}~,\qquad
\partial_{--} A'_{s} &= - \partial_{++} B'_{s-2}~.
\eea
As shown in ref.~\cite{Smirnov:2016lqw}, bilocal operators like \eqref{AABB} are free of short-distance (non-derivative) divergences. Moreover, their expectation value is independent of the separation $\sigma - \sigma'$. Consequently, the limit $\zeta' \to \zeta$ in \eqref{composite-superfield} defines a composite superfield operator that is unique up to total derivative terms. In particular, the integrated operator
\bea
\int\rd^2\s\,\rd\vartheta^+\,\cO_{-}(\zeta)
\label{susy-TTbar}
\eea
is well-defined and manifestly preserves supersymmetry.

We also notice that the operator above is identical to the usual $T\bar{T}$ deformation only 
up to terms that vanish upon using the supercurrent conservation equations that hold on-shell. 
This however is of no consequence when we consider expectation values of \eqref{susy-TTbar} in arbitrary states, since the equations of motion are valid inside correlation functions up to contact terms. In more detail, the relevant descendant of \eqref{composite-superfield} reads
\begin{equation}
\begin{aligned}
(\cD_++\cD'_+)\cO_{-}(\zeta,\zeta')
=\ &
\cT_{----}(\zeta)\,\cT_{++++}(\zeta')
-\cT(\zeta)\,\cT(\zeta')
\\
&
+[\cD_+\cT_{----}(\zeta)-\ri\pa_{--}\cJ_{-}(\zeta)]\,\cJ_{+++}(\zeta')
\\
&
-\ri\cJ_{-}(\zeta)[\pa'_{--}\cJ_{+++}(\zeta')+\pa'_{++}\cJ_-(\zeta')]
\\
&
+\ri(\pa_{--}+\pa'_{--})[\cJ_{-}(\zeta)\,\cJ_{+++}(\zeta')]
\\
&
+\ri(\pa_{++}+\pa'_{++})[\cJ_-(\zeta)\,\cJ_-(\zeta')]~.
\end{aligned}
\end{equation}
When we take its expectation value, Ward identities imply that the second and third line vanish up to contact terms of the form $\delta^2(\sigma - \sigma')$. However, since the expectation value is independent of the separation $\sigma - \sigma'$, these contact terms do not contribute. The third and fourth lines are total derivatives and do not contribute either in states with well-defined energy and momentum.

\subsection{The \texorpdfstring{$\cN=(1,1)$}{N=(1,1)} case}
The $\cN=(1,1)$ case is an obvious generalisation of the previous case. 
We refer the reader to \cite{Gates:1985vk} for 2D $\cN=(1,1)$ off-shell supergravity and its description
in terms of unconstrained prepotentials.
Compared to the $(1,0)$ case of the previous subsections,
superspace is now parametrised by an additional fermionic coordinate, so that 
$\zeta^M = (\s^{++},\s^{--},\vartheta^+,\vartheta^-)$.
 The covariant derivatives and supercharges are defined as
\begin{equation}
\begin{aligned}
\cD_+ & =  \frac{\pa}{\pa \vartheta^+} - {\rm i}  \vartheta^+ \pa_{++}
\,,\qquad&
\cQ_+ 
& = {\rm i}\, \frac{\pa}{\pa \vartheta^+} -   \vartheta^+ \pa_{++}
~,\\
\cD_- & =  \frac{\pa}{\pa \vartheta^-} -{\rm i}  \vartheta^- \pa_{--}
\,,\qquad&
\cQ_- 
& = {\rm i}\, \frac{\pa}{\pa \vartheta^-} -   \vartheta^- \pa_{--}
~,
\end{aligned}
\end{equation}
and the anticommutators read
\begin{equation}
\begin{gathered}
\left\{\cD_+,\cD_+\right\}  = -2\ri\partial_{++}
,\qquad\quad
\left\{\cQ_+,\cQ_+\right\}  = -2\ri\partial_{++}
~,\\
\left\{\cD_-,\cD_-\right\}  = -2\ri\partial_{--}
~,\qquad\quad
\left\{\cQ_-,\cQ_-\right\} = -2\ri\partial_{--}
~,
\\
\left\{\cD_+,\cD_-\right\}  = \left\{\cD_\pm,\cQ_\pm\right\}=\left\{\cQ_+,\cQ_-\right\}  = 0
~.
\end{gathered}
\end{equation}
For a Lorentz  invariant Lagrangian matter system,
$\cN=(1,1)$ supersymmetry implies the existence of two pairs of superfields, $(\cJ_{+++}(\z),\cJ_{-}(\z))$ and 
$(\cJ_{---}(\z),\cJ_{+}(\z))$, which encode the supercurrents and the stress-energy tensor.
These two pairs describe respectively a $(1,0)$ and $(0,1)$ supercurrent multiplet.
By assuming that our matter system can be minimally coupled 
to off-shell $(1,1)$ conformal supergravity together with an unconstrained scalar compensator, 
it can be shown that $\cJ_\pm(\z)$ are expressed in terms of a real scalar current $\cJ(\z)$ as
\bea
\cJ_+(\z)=-\ri\cD_+\cJ(\z)~,~~~~~~
\cJ_-(\z)=\ri\cD_-\cJ(\z)
~.
\eea
In fact, the $\cN=(1,1)$ unconstrained prepotential superfields are:
\begin{equation}
\cH_{+}{}^{--}(\zeta)=-\hf\cH_{+++}(\zeta)\,,\qquad
\cH_{-}{}^{++}(\zeta)=-\hf\cH_{---}(\zeta)\,,\qquad\text{and}\quad \cS(\zeta)\,,
\end{equation}
and possess the following linearised supergravity transformation rules (see \cite{Gates:1985vk} for more detail)
\begin{equation}
\begin{gathered}
\d \cH_{+++} =-\cD_+\cK_{++}~,\qquad\qquad
\d \cH_{---}\ =\ -\cD_{-}\cK_{--}~,\\
\d \cS=\hf(\pa_{++} \cK_{--}+\pa_{--} \cK_{++})+\cK_D
\,.
\end{gathered}
\label{lin-sugra-11}
\end{equation}
Here $\cK_{\pm\pm}(\zeta)$ and $\cK_D(\zeta)$  
are unconstrained and parametrise linearised $(1,1)$ superdiffeomorphisms and
super-Weyl transformations, respectively. 
The supergravity prepotentials comprise various pure gauge components together with the following fields
\begin{equation}
\begin{aligned}
h_{++++}(\s)\equiv& \ -\ri\cD_+\cH_{+++}|_{\vartheta^{\pm}}~,
&\qquad
\psi_{---}&\equiv\ -\cD_+\cD_- \cH_{---}|_{\vartheta^{\pm}}
\,,
\\
h_{----}(\s)\equiv&\ -\ri\cD_-\cH_{---}|_{\vartheta^{\pm}}~,
&\qquad
\psi_{+++}&\equiv\ \phantom{+}\cD_+\cD_- \cH_{+++}|_{\vartheta^{\pm}}
\,,
\\
h(\s)\equiv&\ \phantom{+}\cS|_{\vartheta^{\pm}}
~,&\qquad
\psi_\pm&\equiv\ -\ri\cD_\pm\cS|_{\vartheta^{\pm}}
~,~~~
\,,
\end{aligned}
\end{equation}
where $h_{++++}$ and $h_{----}$ are the traceless components of the linearised metric
while $h$ is its trace, $\psi_{\pm\pm\pm}$ and $\psi_{\pm}$ are the gravitini.

Now consider a general Lorentz invariant matter system coupled to $\mathcal{N}=(1,1)$ supergravity. 
Its action expanded to first order in the supergravity prepotential is
\begin{equation}
S=\frac{\ri}{8}\int\rd^2\s\,\rd\vartheta^+\,\rd\vartheta^-
\Big{[} \cH_{+++}\,\cJ_{---}-\cH_{---}\,\cJ_{+++}+2\cS\,\cJ\Big{]}\,.
\label{action-sources-11}
\end{equation}
Assuming that the equations of motion for the matter are satisfied,
by imposing the invariance of the previous action under 
arbitrary linearised $(1,1)$ superdiffeomorphisms transformations \eqref{lin-sugra-11}
it is straightforward to obtain 
the following conservation equations for the associated  $(1,1)$ supercurrent multiplet
\begin{equation}
\label{supercurrents-11}
\cD_+\cJ_{---}\ =\ \pa_{--}\cJ
~,~~~~~~
\cD_-\cJ_{+++}\ =\ -\pa_{++}\cJ
\,,
\end{equation}
which imply
\begin{align}
	\partial_{--}\cJ_{+++}  = - \partial_{++}\cJ_{-}~,~~~~~~
	\partial_{++}\cJ_{---}  = - \partial_{--}\cJ_{+}~.
\end{align}
If the matter system is superconformal,
the invariance of the action \eqref{action-sources-11} under arbitrary super-Weyl 
implies the extra condition
\bea
\cJ(\zeta)=0
\,.
\eea
It is clear that $\cJ_{\pm\pm\pm}$ and $\cJ$
belong to a supercurrent multiplet where $\cJ$ plays the role of the supertrace.
The components of the stress-energy tensor can be defined as
\begin{equation}
\begin{gathered}
T_{++++}(\s):=\cD_+ \cJ_{+++}|_{\vartheta^{\pm}=0}  
~,\qquad\qquad
T_{----}(\s):=\cD_- \cJ_{---}|_{\vartheta^{\pm}=0} 
~,
\\
 \Theta(\s)= \cD_+ \cJ_{-}|_{\vartheta^{\pm}=0}=\cD_- \cJ_{+}|_{\vartheta^{\pm}=0}
=\ri\cD_+\cD_- \cJ|_{\vartheta^{\pm}=0}
~,
\end{gathered}
\end{equation}
and, due to \eqref{supercurrents-11},
they satisfy
\begin{align}
	\partial_{--}T_{++++}(\s)  = - \partial_{++}\Theta(\s)~,~~~~~~
	\partial_{++}T_{---}(\s)  = - \partial_{--}\Theta(\s)~.
\end{align}

Following the same arguments of the $\cN=(1,0)$ case, it is easy to prove that the composite superfield
\begin{equation}
	\cO(\zeta) \equiv \cJ_{---}(\zeta)\cJ_{+++}(\zeta)-\cJ_{+}(\zeta)\cJ_{-}(\zeta)
	\label{primary-11}
\end{equation}
is well-defined up to terms that vanish upon using the conservation equations and up to total derivative terms. 
Moreover, it is straightforward to show that
\begin{equation}
\cD_+\cD_- \cO|_{\vartheta^{\pm} = 0} = T_{++++}T_{----} - \Theta^2 + \textrm{EOM's} + \textrm{total derivatives}~.
\end{equation}
This shows that the $T\bar{T}$ operator can be constructed as a supersymmetric descendant of an 
$\mathcal{N}=(1,1)$ supersymmetric multiplet, so it preserves the full $\mathcal{N}=(1,1)$ supersymmetry algebra.

Just like in the $\mathcal{N}=(1,0)$ case, it is interesting to write down the deformed action, with deformation 
parameter~$\alpha$, in terms of the original action $S_0\equiv S$ of eq.~\eqref{action-sources-11} and of the 
operator $O(\sigma)$ defined above. At first order in $\a$ we obtain:
\begin{equation}
\begin{aligned}
S_{\alpha}=\ & S_0-\alpha\,\int \de^2\sigma \, O(\sigma)
+\cdots
\\
=\ &\frac{\ri}{8} \int\de^2\sigma\de\vartheta^+\de\vartheta^-\Big[
\big(\mathcal{H}_{+++}-4\ri\alpha\mathcal{J}_{+++}\big)\mathcal{J}_{---}
\\
&\qquad\qquad\qquad\quad~~
-\big(\mathcal{H}_{---}-4\ri\alpha\mathcal{J}_{---}\big)\mathcal{J}_{+++}
\\
&\qquad\qquad\qquad\quad~~
+2\big(\mathcal{S}+4\ri\alpha\mathcal{D}_{+}\mathcal{D}_{-}\mathcal{J}\big)\,\mathcal{J}
\Big]
+\cdots
\,,
\end{aligned}
\end{equation}
much like in the $\mathcal{N}=(1,0)$ case.

\subsection{Energy levels of the deformed theory}
One of the most important features of $T\bar{T}$ deformations is that the energy levels of the deformed theory are related to that of the original theory by an ordinary differential equation~\cite{Zamolodchikov:2004ce}. For a state of energy $H_n$ and vanishing momentum $P_n=0$, this takes the very simple form of eq.~\eqref{eq:TTbargeneral}. The derivation of that formula hinges on the special properties of the (Hilbert) stress-energy tensor, out of which the deforming operator is constructed by means of a point-splitting regularisation procedure.
Since in the super-symmetric set-up we deform the theory by a slightly different operator, we might worry that the $T\bar{T}$ flow of the energy level takes a different form. It is important to note that, as we discussed in the two previous subsections---see in particular the discussion below eq.~\eqref{susy-TTbar}---our point splitting procedure differs from the one of the ``Hilbert'' case only by terms that vanish on-shell. As a result, the two regularised operators are identical up to contact terms. When taking the expectation value of the operator on a state $|n\rangle$ of definite energy $H_n$ and momentum $P_n$, these contact terms give no contributions, because expectation values are independent from the separation of point-split operators~\cite{Zamolodchikov:2004ce}. This means that in the supersymmetric setup we still have the very same ordinary differential equation for the spectrum of a deformed theory.

While the differential equation for $H_n(R,\alpha)$ is the same, its solutions are qualitatively different in the supersymmetric case, due to different initial conditions at~$\alpha=0$. To illustrate this point, it is sufficient to consider the simple case of eq.~\eqref{eq:TTbargeneral}, \textit{i.e.}\ $P_n=0$, following refs.~\cite{Smirnov:2016lqw,Cavaglia:2016oda}. Let us assume that the original, undeformed theory has energy levels given by
\begin{equation}
\label{eq:initial-cond}
H_n(R,\alpha)\Big|_{\alpha=0} = \frac{\Delta_n +\tilde{\Delta}_n - c/12}{R}\,.
\end{equation}
This is the case if the original theory is a (unitary) CFT, in which case $\Delta_n\geq0$ and $\tilde{\Delta}_n\geq0$ are the eigenvalues of the left- and right- $sl(2,\mathbb{R})$ Cartan operators, and $c\geq 0$ is the central charge. More generally, we expect this to be approximately correct for $R$ small enough, in which case $c$ is the central charge of a suitable UV CFT. It is easy to solve eq.~\eqref{eq:TTbargeneral} with initial conditions~\eqref{eq:initial-cond}:
\begin{equation}
\label{eq:energylavels}
H_n(R,\alpha) = \frac{-R + \sqrt{R^2 + 4 \alpha \Big(\Delta_n +\tilde{\Delta}_n - c/12 \Big)}}{2\alpha}\,.
\end{equation}
Note that when $c>0$ and $|\alpha|$ is large enough, there are always cases where the square-root becomes imaginary: for $\alpha<0$ this happens whenever $\Delta_n + \tilde{\Delta}_n >c/12$, which is the case for infinitely many excited states; for $\alpha>0$ this happens at least for the ground state $|0\rangle$ where $\Delta_0=\tilde{\Delta}_0=0$ (as well as $P_0=0$). The former case has been given the interpretation of a ``holographic cutoff''~\cite{McGough:2016lol}, while the latter at least in the simplest cases can be understood as the tachyon of bosonic string theory~\cite{Cavaglia:2016oda}.%
\footnote{More generally, this behaviour arises when putting a $T\bar{T}$ deformed theory on the torus, as commented on in \textit{e.g.}\ ref.~\cite{Aharony:2018bad}.}
We see now that supersymmetric theories are very special in this regard, because for suitable boundary conditions of the fermions the vacuum energy in finite volume is zero. Therefore, $c=0$ in eq.~\eqref{eq:energylavels} so that we have a regular $T\bar{T}$ flow \textit{for all states} for $\alpha>0$. Moreover, the supersymmetric ground state is protected under the $T\bar{T}$ flow, having $H_0(R,\alpha)=0$ for all~$\alpha\in\mathbb{R}$. Note however that even in the supersymmetric case a deformation with $\alpha<0$ would generically lead to complex energy levels for the excited states, in good accord with the ``holographic cutoff'' interpretation of ref.~\cite{McGough:2016lol}.


\section{Example:  \texorpdfstring{$\mathcal{N}=(1,0)$}{N=(1,0)} \texorpdfstring{$T\bar{T}$}{TTbar}-deformed action}
\label{sec:TTbarN10}
In order to illustrate the ideas discussed above, let us consider in some detail the simplest supersymmetric setup, 
which is a two-dimensional theory with $\mathcal{N}=(1,0)$ supersymmetry. The free action for 
a $\cN=(1,0)$ real scalar multiplet is
\begin{equation}
\label{eq:free10action}
S_0=\int \de^2\sigma \Big[\frac{1}{2}\partial_{++}X\partial_{--}X+\frac{\ri}{2}\,\psi_{+}\partial_{--}\psi_+\Big]\,.
\end{equation}
This is invariant under the supersymmetry transformations
\begin{equation}
\label{eq:susyvar}
\delta_Q X=-\ri\, \epsilon_-\psi_+\,,
\qquad
\delta_Q \psi_+ = \epsilon_{-}\partial_{++}X\,.
\end{equation}
It is important to note that the previous supersymmetry transformations close \emph{off-shell}. This is a 
feature that  the $(1,0)$ case makes particularly simple compared to higher supersymmetric cases
where one might have to introduce auxiliary fields to close supersymmetry off-shell.
The previous real scalar multiplet is equivalently described in $\cN=(1,0)$ superspace by a real unconstrained 
superfield $\varphi(\s,\vartheta^+)$ such that
\bea
\varphi(\s,\vartheta^+)
=
X(\s)
-\ri\vartheta^+\psi_+(\s)
~.
\eea
The free action \eqref{eq:free10action} is equivalently written in superspace as
\begin{equation}
\label{eq:free10action-superspace}
S_0=
\hf\int\rd^2\s\rd\vartheta^+\cD_+\varphi\,\ri\pa_{--}\varphi
\,.
\end{equation}

We want to consider a $T\bar{T}$ deformed action $S_\a$ with parameter~$\alpha$ such that
\begin{equation}
\label{eq:TTbardiffeq}
\partial_\alpha S_{\alpha} =- \text{det}[T],\qquad \lim_{\alpha\to0}S_{\alpha}=S_0\,,
\end{equation}
where the stress-energy tensor is computed out of $S_\alpha$ itself.
There are several ways to do this. For instance, one could construct the deformation order by order in $\alpha$, 
following ref.~\cite{Cavaglia:2016oda}. At each given order, the stress-energy tensor will be a polynomial in 
$\partial_{\pm\pm}X$ and $\psi_{+}\partial_{\pm\pm}\psi_{+}$. This, together with the Grassmannian 
nature of~$\psi_{+}$, leads to the simple ansatz
\begin{equation}
\label{eq:ansatz10}
S_{\alpha} = \int\de^2\sigma \Big[\frac{1}{\alpha}A(\var) + \ri\,B(\var)\,\psi_{+}\partial_{--}\psi_+ 
+ \ri\,\alpha\,C(\var)\,\big(\partial_{--}X\big)^2\psi_{+}\partial_{++}\psi_+\Big],
\end{equation}
where we have accounted for the fact that $\alpha$ has dimension $[\alpha]=(-1,-1)$ and introduced the dimensionless combination
\begin{equation}
\var = \alpha\, \partial_{++}X\partial_{--}X\,.
\end{equation}
Imposing \eqref{eq:TTbardiffeq} in terms of the Noether stress-energy tensors then leads to a system of ordinary 
differential equations for $A(\var)$, $B(\var)$ and $C(\var)$, much like in ref.~\cite{Bonelli:2018kik}. Relegating the 
details to appendix~\ref{app:onecommazeroaction} we have
\begin{equation}
\begin{aligned}
0=&A(\var)\big[1+A(\var)\big]-\var\,A'(\var)\,\big[1+2A(\var)\big],\\
0=&\var\,A'(\var)\big[B(\var)+\var\,C(\var)\big] +\var\,B'(\var)-A(\var)\big[B(\var)-2\var\,B'(\var)\big],\\
0=&A'(\var)\,B(\var)+C(\var)\,\big[1+A(\var)+\var\,A'(\var)\big]+\var\,C'(\var)\,\big[1+2A(\var)\big],
\end{aligned}
\end{equation}
which we want to solve with initial conditions at small~$\var$
\begin{equation}
A(z)=\frac{1}{2}x+O(\var^2)\,,\quad
B(z)=\frac{1}{2}+O(\var^1)\,,\quad
C(z)=0+O(\var^0)\,.
\end{equation}
The differential equation for $A(\var)$ is solved by a bosonic action of Nambu-Goto form; solving the remaining 
equations we get
\begin{equation}
\label{eq:solNoether}
S_\alpha:\qquad A(\var)= \frac{\sqrt{1+2\var}-1}{2}\,,\quad
B(\var)=\frac{1}{4}+\frac{1+\var}{4\sqrt{1+2\var}}\,,\quad
C(\var)=-\frac{1}{4\sqrt{1+2\var}}\,.
\end{equation}
It is easy to verify that, using the supersymmetry variation~\eqref{eq:susyvar},
\begin{equation}
\delta_Q\,S_{\alpha}\neq 0\,,
\end{equation}
whenever~$\alpha\neq0$. In fact, it is not difficult to construct the unique Lagrangian with bosonic 
part~$\mathcal{L}_{\text{bos},\alpha}=A(\var)/\alpha$ which is invariant under~\eqref{eq:susyvar}. 
As we show in appendix~\ref{app:onecommazeroaction}, imposing supersymmetry leads to a very simple 
differential equation,
\begin{equation}
B(\var)=A'(\var),\qquad 2C(\var)+\var\,C'(\var)=B'(\var)\,,
\end{equation}
which for our bosonic Lagrangian $A(\var)$ is solved by
\begin{equation}
\label{eq:solSusy}
S_{\text{susy}}: 
\qquad
B(\var)=\frac{1}{2\sqrt{1+2\var}}\,,\qquad
C(\var)=\frac{1}{2\var^2}-\frac{1+\var}{2\var^2\sqrt{1+2\var}}\,.
\end{equation}
It is also simple to show that in superspace the resulting manifestly supersymmetric action takes the form
\bea
S_{\rm susy} 
=
\int\rd^2\s\rd\vartheta^+\,
\frac{\sqrt{1+2\cX}-1}{2\cX}
\,
\cD_+\varphi\,\ri\pa_{--}\varphi
~,~~~~~~
\cX\equiv\a\,\pa_{++}\varphi\,\pa_{--}\varphi
~.
\eea
This can actually be proven to satisfy exactly a $T\bar T$ flow driven by the primary operator \eqref{susy-TTbar-10}.
Such an action can also be straightforwardly generalised to a 
manifestly $\mathcal{N}=(1,1)$ supersymmetric  action:\footnote{The action \eqref{Ssusy11} satisfies the $T\bar{T}$ flow equation
up to equations of motion. More precisely, the exact off-shell solution of the flow driven by the primary operator
\eqref{primary-11} has an explicit dependence also on the combination $\mathcal{Y}:=(\cD_+\cD_-\F)^2$. This modify eq.~\eqref{Ssusy11} by appearing in front of $\cD_+\F\cD_-\F$.
However, for both the off-shell $T\bar{T}$ flow action and
\eqref{Ssusy11} the equations of motion imply $\mathcal{Y}\,\cD_+\F\cD_-\F=0$ so that any dependence such on $\mathcal{Y}$ can be neglected on~shell.}
\bea
S_{\rm susy} 
=
\int\rd^2\s\rd\vartheta^+\rd\vartheta^-\,
\frac{\sqrt{1+2\cX}-1}{2\cX}
\,
\cD_+\Phi\,\cD_-\Phi
~,~~~~~~
\cX\equiv\a\,\pa_{++}\Phi\,\pa_{--}\Phi
~,
\label{Ssusy11}
\eea
where $\Phi$ is an unconstrained real superfield describing the off-shell $\mathcal{N}=(1,1)$ scalar multiplet, 
$\Phi
=X
-\ri\vartheta^+\psi_+
-\ri\vartheta^-\psi_-
+\ri\vartheta^+\vartheta^-F$. Notice that this depends on the real auxiliary field $F(\s)$. 
We will consider in more detail $\mathcal{N}=(1,1)$ theories in the next section restricting to the case where 
$F(\s)$ is integrated out and $\cN=(1,1)$ supersymmetry is realised on-shell.

Let us comment on our results.
Note that, the difference~$\Delta$ between the two actions, $S_\a$ and $S_{\rm susy}$,
amounts to the difference between the $T\bar{T}$ operator constructed out of the Noether tensor and the supersymmetric descendant of the operator $O_-$ considered in section~\ref{sec:TTbarsusy}, \textit{cf.}\ eq.~\eqref{susy-TTbar-10}. Hence we expect $\Delta$ to vanish on-shell, which is easy to verify. Indeed, the equations of motion for the fermion~$\psi_+$ are identical for $S_\alpha$ and $S_{\text{susy}}$, essentially due to the ratio $C(\var)/B(\var)$ being the same in the two cases, and take the form
\begin{equation}
\label{eq:fermioneom}
\partial_{--}\psi_+=\frac{\alpha}{1+\var+\sqrt{1+2\var}}(\partial_{--}X)^2 \partial_{++}\psi_+
\,.
\end{equation}
From this fact it immediately follows that the fermionic pieces of \textit{both}~$S_\alpha$ and $S_{\text{susy}}$ vanish when imposing \textit{the same} equation of motion for~$\psi_+$. Hence, the difference $\Delta$ vanishes when we impose eq.~\eqref{eq:fermioneom}.

One might wonder whether our conclusion is special to the Noether stress-energy tensor, or whether it holds more generally. We start by observing that for the case at hand the Hilbert stress-energy tensor takes the same form as the Noether one; technically, this is because the spin-connection drops out for the real fermion~$\psi_+$. Nonetheless, it is still possible to define an \textit{improved stress-energy tensor}, \textit{i.e.}\ one that is manifestly symmetric (and, for the free theory, traceless). In that case too we find that the resulting $T\bar{T}$-action can be expressed in terms of coefficients $A(\var)$, $B(\var)$, $C(\var)$ as
\begin{equation}
\label{eq:improved}
S_{\alpha}^{\text{impr}}:\quad A(\var)=\frac{\sqrt{1+2\var}-1}{2},\quad
B(\var)=\frac{1+\sqrt{1+2\var}}{4\,\sqrt[4]{1+2\var}},\quad
C(\var)=\frac{1-\sqrt{1+2\var}}{4\var\,\sqrt[4]{1+2\var}}.
\end{equation}
Despite the rather different form, this action once again yields the same equation of motion for~$\psi_+$, so that again the deformations are equivalent on-shell.

Note that, though only $S^{\rm susy}$ is invariant under the 
supersymmetry transformations \textit{of the free theory}, eq.~\eqref{eq:susyvar},
both $S_{\alpha}$ and $S_{\alpha}^{\text{impr}}$ can be shown to be invariant under a \textit{deformed}
set of supersymmetry transformations. This is consistent with the results of the previous section that point out 
that only deformations based on $O_-$ will manifestly preserve supersymmetry while other deformations
will do so only up to terms vanishing on-shell.

\section{\texorpdfstring{$T\bar{T}$}{TTbar} deformations and superstring theory}
\label{sec:TTbarstring}
While in the case of a {free} $\mathcal{N}=(1,0)$ theory we could explicitly construct a $T\bar{T}$ deformation for finite~$\alpha$ (following the strategy of ref.~\cite{Bonelli:2018kik}), doing so for a generic theory becomes rather cumbersome. We will see in this section how to exploit a link with strings in light-cone gauge, first highlighted in ref.~\cite{Baggio:2018gct}, to construct the $T\bar{T}$ deformation of more general theories; in particular, we will construct the $T\bar{T}$ deformation of the free theory of eight  on-shell $\mathcal{N}=(1,1)$ scalar multiplets,
 which emerges from superstrings on~$\mathbb{R}^{1,9}$.

\subsection{Uniform light-cone gauge and \texorpdfstring{$T\bar{T}$}{TTbar} deformations}
\label{sec:lcgauge}
To begin with, let us briefly review the link between strings and $T\bar{T}$ deformations. The first observation in this sense was that a relation exists between the $T\bar{T}$-deformed Lagrangian 
for free bosons and the Nambu-Goto action in flat space~\cite{Cavaglia:2016oda}.
As described in ref.~\cite{Baggio:2018gct}, this relation becomes particularly transparent in the uniform light-cone gauge of refs.~\cite{Arutyunov:2004yx,Arutyunov:2005hd,Arutyunov:2006gs}, see also the review~\cite{Arutyunov:2009ga}. Let us briefly review that argument.

Consider a string background with shift isometries along the time coordinate $t$ and along the spacial coordinate $\varphi$.%
\footnote{We take the signature of the target space metric to be negative for~$t$ and positive for all space coordinates.}
 We introduce the one-parameter family of light-cone coordinates%
\footnote{We indicate the target-space light-cone coordinates with indices~$\pm$, which are raised and lowered with the target-space metric described in appendix~\ref{app:conventions}. These should not be confused with the two-dimensional superspace indices used in the previous sections in expressions like $\partial_{\pm\pm}$, $\cO_{-}$, $\cJ_{+++}$, \textit{et caetera}.}
\begin{equation}
\label{eq:lccoords}
X^{+} = a\,\varphi+(1-a)\,t\,,\qquad
X^{-} = \varphi-t\,,\qquad 0\leq a\leq 1\,.
\end{equation}
The most notable cases are given by $a=0$ and $a=1$, where $X^+$ becomes aligned to the temporal and spatial coordinate, respectively, and the light-like choice~$a=\tfrac{1}{2}$; however we can consider any $0\leq a \leq 1$. Introducing the momenta canonically conjugate to~$t$ and~$\varphi$, denoted by~$P_t$ and $P_\varphi$ respectively, we find
\begin{equation}
P_+=P_\varphi+P_t\,,\qquad
P_-=(1-a)\,P_\varphi-a\,P_t\,.
\end{equation}
It is convenient to relate this to physical quantities by introducing the Noether charges $E$ and $J$ corresponding to shifts under~$t$ and~$\varphi$. Then we have
\begin{equation}
\label{eq:conjugatecharges}
\int\limits_0^R\de\sigma^1 P_+= J-E,\qquad
\int\limits_0^R\de\sigma^1 P_-= J+a\,(E-J)\,,
\end{equation}
where $R$ is the size of the closed-string worldsheet.

The \textit{uniform light-cone gauge}~\cite{Arutyunov:2004yx,Arutyunov:2005hd,Arutyunov:2006gs} is fixed by imposing
\begin{equation}
\label{eq:lcgauge}
X^+=\sigma^0\,,\qquad P_-=1\,,
\end{equation}
where~$\sigma^0$ is the worldsheet time.
The name \textit{uniform} emphasises that the momentum density~$P_-$ is constant on the string. Equivalently, this can be seen as fixing the T-dual coordinate to~$X^-$ to be~$\widetilde{X}^{-}=\sigma^1$~\cite{Kruczenski:2004cn} (see also appendix~\ref{app:lcgauge}).
From this rather simple condition, a number of remarkable facts follow. 
Firstly, since $X^+$ is identified with the worldsheet time~$\sigma^0$, the Hamiltonian~$H$ of the two-dimensional model is identified with the conjugate momentum~$P_+$; in particular
\begin{equation}
H = E-J\,.
\end{equation}
Secondly, the integral of~$P_-$ in eq.~\eqref{eq:conjugatecharges} can be immediately done, and it fixes the worldsheet size~$R$ in terms of the charges~$E$ and~$J$ and the parameter~$a$:
\begin{equation}
\label{eq:Rofa}
R=R(a)= J+a\,H\,.
\end{equation}
In other words, for any given $a$ the size of the worldsheet is \textit{state-dependent} and fixed in terms of each state's energy~$E$ and charge~$J$---reminiscent of what happens to~$T\bar{T}$-deformed theories. We can be more quantitative by looking more closely to the charge~$H$. This is given by the integral of the $a$-dependent%
\footnote{The explicit form of $P_+(a)$ can be found quite straightforwardly, at least for the bosonic part of the theory, see \textit{e.g.}\ ref.~\cite{Arutyunov:2009ga}; we will analyse this in more detail for superstrings in the next section.}
Hamiltonian density $P_+=P_+(a)$ over the volume~$R(a)$. On the other hand, by gauge invariance the physical spectrum of $H$ cannot depend on $a$, so that
\begin{equation}
\frac{\de}{\de a}H=0=\frac{\de}{\de a}\int\limits_0^{R(a)}\de\sigma^1 P_+(a)\,.
\end{equation}
That is to say: tuning $a$ changes the Hamiltonian density in such a way as to precisely compensate the rescaling of~$R(a)$. Given that the $a$-dependence of $R$ is precisely that of a $T\bar{T}$ deformation for a state without worldsheet momentum,%
\footnote{%
This is not really a restriction, as in string theory the level-matching constraint dictates that the worldsheet momentum of physical states should vanish up to winding. Non-trivial winding sectors can also be incorporated, see ref.~\cite{Dei:2018mfl}.
} tuning $a$ mimics a $T\bar{T}$ deformation of the Lagrangian.

Another way to understand how this similarity arises is by thinking of $T\bar{T}$ deformations as CDD factors~\cite{Castillejo:1955ed}, as in refs.~\cite{Smirnov:2016lqw, Cavaglia:2016oda}. Under a shift of the gauge parameter~$a$ the worldsheet S~matrix is modified by a CDD factor~$e^{i\,a\, \Phi_{\text{CDD}}}$ of the form~\cite{Arutyunov:2009ga}
\begin{equation}
\Phi_{\text{CDD}}(p_i,p_j) = H_i\, p_j - H_j\,p_i\,,
\end{equation}
so that the spectrum is unchanged; schematically, the Bethe-Yang equations take the form
\begin{equation}
e^{i p_i (J+a\,H)}\prod_j e^{i\,a\,\Phi_{\text{CDD}}(p_i,p_j)}\,S(p_i,p_j)\big|_{a=0}=1\,,
\end{equation}
which is independent of~$a$ when we use that $\sum_j p_j=0$ and $\sum_j H_j=H$. The form of such a CDD factor is indeed what expected for a $T\bar{T}$ deformation~\cite{Smirnov:2016lqw, Cavaglia:2016oda} and what observed for flat-space strings~\cite{Dubovsky:2012wk}.
This highlights the generality of the relation between $T\bar{T}$-deformed Lagrangians and light-cone gauge fixed strings.

It is worth stressing once more that, unlike what happens when performing a $T\bar{T}$ deformation, changing the gauge parameter $a$ \textit{does not} change the spectrum. What does change the spectrum is to change the parameter~$a$ \textit{in the Lagrangian density}, \textit{while keeping the volume $R$ fixed}---or vice-versa.

\subsection{Superstrings in flat space as a \texorpdfstring{$T\bar{T}$}{TTbar} deformation}
\label{sec:n88}

Let us now apply the logic illustrated above to superstring in flat space. We will explicitly see that the gauge-fixed (and $\kappa$-gauge-fixed) Lagrangian in uniform light-cone gauge is related to the $T\bar{T}$ deformation of a free theory of eight  on-shell $\mathcal{N}=(1,1)$ scalar multiplets. The deformation parameter~$\alpha$ is related to the gauge parameter~$a$ by
\begin{equation}
\label{eq:alphaa}
\alpha = a-\frac{1}{2}\,.
\end{equation}
Indeed the Lagrangian is free at $a=1/2$.
This theory also exhibits an $\mathfrak{so}(8)$ flavour symmetry, which is the manifest isometry of $\mathbb{R}^{1,9}$ that survives in light-cone gauge.

The derivation of the uniform light-cone gauge-action including fermions is relatively straightforward. We follow the procedure outlined in ref.~\cite{Arutyunov:2014jfa}; for the sake of completeness, we include some intermediate steps and technical details in appendix~\ref{app:GS}. We start from the Green-Schwarz action
\begin{equation}
\label{eq:GSaction}
S=-\frac{1}{2}\int\de^2\sigma \Big[\gamma^{ab}\eta_{\mu\nu}\Pi_a^\mu\Pi_b^\nu 
+\epsilon^{ab}\big(2\ri\partial_a X^\mu \bar{\theta}^I\Gamma_\mu \sigma^{(3)}_{IJ} \partial_b\theta^J
+2\bar{\theta}^1\Gamma^\mu \partial_a\theta^1\,\bar{\theta}^2\Gamma_\mu \partial_b\theta^2
\big)
\Big],
\end{equation}
with
\begin{equation}
\Pi^\mu_a=\partial_a X^\mu+ \ri\bar{\theta}^I\Gamma^\mu\delta_{IJ} \partial_a \theta^{J}.
\end{equation}
Some definitions are in order. The worldsheet metric \textit{with unit-determinant} is $\gamma^{ab}$; the 10-dimensional flat target-space metric is $\eta^{\mu\nu}$, with signature $(-,+\dots+)$.
The target-space indices are $\mu=0,1,\dots 9$ while the worldsheet ones are $a=0,1$.
We denote the two 10-dimensional Majorana-Weyl spinors as $\theta^{I=1,2}$, while $\Gamma^\mu$ are 10-dimensional Gamma matrices; we define $\bar{\theta}^{I}=(\theta^I)^{\dagger}\Gamma^0$. Further details on our conventions are collected in appendix~\ref{app:GS}.

We can now fix uniform light-cone gauge like in eqs.~(\ref{eq:lccoords}--\ref{eq:lcgauge}), with the identification $t\equiv X^0$ and $\varphi\equiv X^9$ for the light-cone coordinates. Additionally, we must fix the $\kappa$ gauge; this can be done by setting
\begin{equation}
\label{eq:kappagauge}
\big(\Gamma^9+\Gamma^0\big)\,\theta^I=0,\qquad I=1,2\,,
\end{equation}
which halves the fermionic degrees of freedom. This leaves us with eight real fermions in $\theta^1$, which we denote by $\{\psi_-^j\}_{j=1,\dots8}$, and just as many in $\theta^2$, which we denote by $\{\psi_+^j\}_{j=1,\dots8}$. With some straightforward algebraic manipulations (\textit{cf.}\ appendix~\ref{app:bilinears}) we can show that the fermions enter in the Lagrangian through bilinears of the form $(\theta^I)^{\text{t}}\partial_a\theta^I$. For this reason, let us define the short-hand notations
\begin{equation}
\label{eq:fermionshorthand}
\begin{aligned}
\Psi^1_a \equiv (\theta^1)^{\text{t}}\partial_a\theta^1 
= \frac{1}{4}\sum_{j=1}^8 \psi_{-}^i \partial_a \psi_{-}^i\,,
\qquad
\Psi^2_a \equiv (\theta^2)^{\text{t}}\partial_a\theta^2
= \frac{1}{4}\sum_{j=1}^8 \psi_{+}^i \partial_a \psi_{+}^i\,,
\end{aligned}
\end{equation}
where we have chosen the normalisation of the fields~$\psi_{\pm}$ for later convenience.

A convenient way to fix uniform light-cone gauge is to perform a T-duality along $X^-$~\cite{Kruczenski:2004cn} (see also appendix~\ref{app:Tduality}). Denoting the T-dual coordinate as~$\widetilde{X}^{-}$, we find an action of the form
\begin{equation}
\label{eq:ansatzAct}
S=-\frac{1}{2}\int\de^2\sigma\big(
\gamma^{ab}\mathcal{A}_{ab}+\mathcal{B}
\big)\,,
\end{equation}
with%
\footnote{The worldsheet indices $a,b$ of $\mathcal{A}_{ab}$ are implicitly symmetrised.}
\begin{equation}
\label{eq:Tdualaction}
\begin{aligned}
\mathcal{A}_{ab}=&\,\frac{1}{2\alpha}\big( \partial_a \widetilde{X}^{-}\partial_b \widetilde{X}^{-} - \partial_a X^+\partial_b X^+  \big)+\partial_aX^i\partial_bX^i\\
&\qquad\qquad\quad
+2\ri\Big[\partial_a\widetilde{X}^{-} \big(\Psi_b^{1}-\Psi_b^{2}\big)+\partial_a X^+ \big(\Psi_b^{1}+\Psi_b^{2}\big)\Big]
+ 8\alpha\,\Psi_a^{1}\Psi_b^{2},\\
\mathcal{B}=&\;\epsilon^{ab}\,\Big[-\frac{1}{\alpha}\partial_aX^+\partial_b\widetilde{X}^{-} \\
&\qquad\qquad\quad
+2\ri(\partial_a\widetilde{X}^{-}(\Psi^{1}_b+\Psi^{2}_b)+\partial_a X^+(\Psi^{1}_b-\Psi^{2}_b))-8\alpha\, \Psi^{1}_a\Psi^{2}_b
\Big]\,,
\end{aligned}
\end{equation}
which we want to evaluate in the gauge%
\footnote{%
We have already dropped a term of the form  $\epsilon^{ab}\partial_a\widetilde{X}^-\partial_bX^-\approx\dot{X}^-$ in the expression of $\mathcal{B}$ in eq.~\eqref{eq:Tdualaction} as this is a total (temporal) derivative in the gauge-fixed theory.
}
\begin{equation}
X^+=\sigma^0,\qquad \widetilde{X}^{-}=\sigma^1\,.
\end{equation}
Note that the parameter $\alpha$ is related to the gauge parameter by eq.~\eqref{eq:alphaa}, and the index $i$ runs over the transverse bosons, $i=1,\dots 8$.
Imposing these conditions and eliminating the worldsheet metric $\gamma^{ab}$ by virtue of its equations of motion, we obtain the action
\begin{equation}
\label{eq:NGaction}
S=-\int\de^2\sigma\Big(
\sqrt{-|\mathcal{A}|}+\frac{1}{2}\mathcal{B}
\Big)\,.
\end{equation}
with%
\footnote{%
In the formula below we introduced the short-hand notation
$A^{[i}B^{j]}\equiv A^iB^j-A^jB^i$.}
\begin{equation}
\label{eq:ABfinal}
\begin{aligned}
-|\mathcal{A}|=&\frac{1}{4\alpha^2}- \frac{1}{2\alpha}\Big[\partial_{++}X^i\partial_{--}X^i+2\ri\big(\Psi_{--}^1 + \Psi_{++}^2\big)\Big]\\
&-\frac{1}{4}\partial_{--}X^i\partial_{++}X^j\,\partial_{--}X^{[i}\partial_{++}X^{j]}-2\Psi^{[1}_{--}\Psi^{2]}_{++} -\big(\Psi^1_{--} + \Psi^2_{++}\big)^2\\
&-\ri\Big[\big(\partial_{--}X^i\big)^2\Psi^1_{++} + (\partial_{++}X^i)^2\Psi^2_{--} - \partial_{++}X^i\partial_{--}X^i\big(\Psi^{1}_{--} + \Psi^2_{++}\big)\Big]\\
&-2\alpha\Big[
\big(\partial_{++}X^i\Psi^1_{--} - \partial_{--}X^i\Psi^1_{++}\big)
\big(\partial_{++}X^i\Psi^2_{--} - \partial_{--}X^i\Psi^2_{++}\big)\\
&\qquad\qquad\qquad\qquad\quad
-2\ri\big(\Psi^1_{--} + \Psi^2_{++}\big)\Psi^{[1}_{--}\Psi^{2]}_{++}
\Big]
+4\alpha^2\,\big(\Psi^{[1}_{++}\Psi^{2]}_{--}\big)^2\,,\\
\mathcal{B}=&-\frac{1}{\alpha}-2\ri\big(\Psi^1_{--} + \Psi^2_{++} \big) -4\alpha\, \Psi^{[1}_{--}\Psi^{2]}_{++}\,.
\end{aligned}
\end{equation}

The expression above is fairly involved. It is instructive to explicitly consider a few limits. First of all, there exists a limit in which the light-cone gauge-fixed Lagrangian for flat superstrings is \textit{free}. This happens when both $X^+$ and $X^-$ are light-like, \textit{i.e.}\ for  $a=\frac{1}{2}$ and $\alpha=0$, \textit{cf.}\ eq.~\eqref{eq:lccoords}. Indeed we find
\begin{equation}
\begin{aligned}
-\sqrt{-|\mathcal{A}|}-\frac{1}{2}\mathcal{B}&\,=
\frac{1}{2}\partial_{++} X^i\partial_{--}X^i +2\ri\big(\Psi_{--}^1+\Psi_{++}^2\big)+O(\alpha)\\
&\,=\frac{1}{2}\partial_{++} X^i\partial_{--}X^i
+\frac{\ri}{2}\psi_+^i\partial_{--}\psi_+^i
+\frac{\ri}{2}\psi_-^i\partial_{++}\psi_-^i +O(\alpha)\,.
\end{aligned}
\end{equation}
Another interesting limit is restricting to a theory of  bosons only. In this case we get the non-linear Lagrangian
\begin{equation}
-\sqrt{-|\mathcal{A}|}-\frac{1}{2}\mathcal{B}=
\frac{1-\sqrt{1-2\alpha\,\partial_{++}X^i\partial_{--}X^i+ \alpha^2\,\partial_{++}X^i\partial_{--}X^j\partial_{++}X^{[j}\partial_{--}X^{i]}}}{2\alpha}
\,.
\end{equation}
This is a $T\bar{T}$ deformation of the free-boson action with parameter~$-\alpha$~\cite{Cavaglia:2016oda}. This indeed fits with our string-theory construction: recall that the worldsheet size in terms of $\alpha$ is $R=R_0+\alpha\,H$, see eq.~\eqref{eq:Rofa}. In this case, the transformation of the Lagrangian is compensating for the transformation of the radius, so in our convention this corresponds to the \textit{opposite} of a canonical $T\bar{T}$ transformation.

We can also consider the $\mathcal{N}=(1,0)$ Lagrangian by restricting to the boson $X\equiv X^1$ and one single chiral fermion $\psi_+\equiv \psi^1_+$. The action again simplifies and we find
\begin{equation}
\begin{aligned}
\begin{aligned}
-|\mathcal{A}|=&\frac{1}{4\alpha^2}- \frac{1}{2\alpha}\Big[\frac{1}{2}\partial_{++}X\partial_{--}X+\frac{\ri}{2}\psi_+\partial_{--}\psi_+\Big]\\
&\qquad\qquad
-\frac{\ri}{4}\big(\partial_{--}X\big)^2\psi_+\partial_{++}\psi_+
+\frac{\ri}{4}\,\partial_{++}X\partial_{--}X\,\psi_{+}\partial_{--}\psi_+\,,\\
\mathcal{B}=&-\frac{1}{\alpha}-\frac{\ri}{2}\psi_+\partial_{--}\psi_+\,,
\end{aligned}
\end{aligned}
\end{equation}
which, upon expanding~$\sqrt{-|\mathcal{A}|}$ over the fermions, precisely reproduces~(\ref{eq:ansatz10}--\ref{eq:solNoether}) up to changing $\alpha\to-\alpha$ as expected.

Finally, it is easy to explicitly verify that the full action \eqref{eq:ABfinal} satisfies the $T\bar{T}$ differential equation~\eqref{eq:TTbardiffeq} by following the approach we outlined in the $\mathcal{N}=(1,0)$ case, see appendix~\ref{app:nother10}. As we have remarked in the $\mathcal{N}=(1,0)$ case, the real fermions $\psi^i$ do not couple to the spin connection so that the Noether and Hilbert stress-energy tensors coincide for this action.
In addition to providing an explicit deformation of a more general free supersymmetric action
associated to $\a=0$, this example shows that the link between string theory and  $T\bar{T}$ deformations holds also when fermionic degrees of freedom are included. The precise relation requires fixing (light-cone) $\kappa$-gauge as well as the bosonic gauge.

\subsection{Deformation from the induced worldsheet metric}
\label{sec:wsmetric}
We have obtained the $T\bar{T}$ deformed Lagrangian of eq.~\eqref{eq:NGaction} by integrating out the worldsheet metric $\gamma_{ab}$ in eq.~\eqref{eq:ansatzAct} (and of course fixing uniform light-cone gauge). This means that we can think of~\eqref{eq:NGaction} as the NLSM action~\eqref{eq:ansatzAct} \textit{on a specific worldsheet metric} given by
\begin{equation}
\gamma_{ab}= \mathcal{A}_{ab}\,,
\end{equation} 
which follows from the equations of motion. This is somewhat reminiscent of ref.~\cite{Conti:2018tca} where it was argued that $T\bar{T}$ deformations can be understood in terms of a field-dependent ``wordlsheet'' metric.%
\footnote{We are grateful to Stefano Negro for bringing this observation to our attention and for very helpful discussions related to this point.
}
 To make this analogy more manifest let us write down explicitly eq.~\eqref{eq:ansatzAct} in uniform light-cone gauge \eqref{eq:lcgauge}. We also restrict to the bosons for simplicity. Then
\begin{equation}
\label{eq:constraintaction}
S=-\frac{1}{2}\int\de^2\sigma\Big[
\frac{1}{2\alpha}\Big(\gamma^{ab}\,\eta_{ab}-2\Big)+\gamma^{ab}\partial_aX^i\partial_bX^i
\Big]\,,
\end{equation}
where the unit-determinant metric~$\gamma_{ab}$ is given by
\begin{equation}
\label{eq:wsmetric}
\gamma_{ab}=\frac{h_{ab}}{\sqrt{-h}},\qquad
h_{ab}=\left(
\begin{array}{cc}
-\displaystyle\frac{1}{\alpha}+2\partial_{0}X^i\partial_{0}X^i&2\partial_{0}X^i\partial_{1}X^i\\
2\partial_{0}X^i\partial_{1}X^i&+\displaystyle\frac{1}{\alpha}+2\partial_{1}X^i\partial_{1}X^i
\end{array}
\right)\,.
\end{equation}
Firstly, observe that in eq.~\eqref{eq:constraintaction} the last term gives the matter action (the transverse fields of the string) minimally coupled to the metric. The first bracket, instead, is not covariant and implements the light-cone gauge constraints. Indeed in the limit $\alpha\to0$ (where we expect a free theory) this forces $\gamma_{ab}=\eta_{ab}$.
This is rather reminiscent both of the construction of ref.~\cite{Dubovsky:2017cnj}, where the $T\bar{T}$ deformation emerges from coupling the matter action to Jackiw-Teitelboim gravity, and of the more recent observation of ref.~\cite{Conti:2018tca} that the $T\bar{T}$ transformation may be ``undone'' by introducing a field-dependent metric.
However, the precise forms both of the action \eqref{eq:constraintaction} and of the metric \eqref{eq:wsmetric} differ from what considered in refs.~\cite{Dubovsky:2017cnj,Conti:2018tca}---despite eventually leading to the same $T\bar{T}$-deformed Lagrangian, at least for the case of free bosons. It would certainly be interesting to explore this connection in more detail and for more general theories. It is worth remarking that, already for the case of free bosons \textit{and} fermions, the form of the metric becomes significantly more complicated, as it can be seen from $\mathcal{A}_{ab}$ in eq.~\eqref{eq:ABfinal}; the same will be true when considering more general geometries. We hope to return to this in the near future.

\section{Conclusions and outlook}
\label{sec:conclusions}
We have seen that the structure of $T\bar{T}$ deformations is compatible with supersymmetry, and can be studied quite explicitly in the case of $\mathcal{N}=(1,0)$ and $\mathcal{N}=(1,1)$ supersymmetry. There, the operator 
$O=\text{det}[T^{ab}]$
is a full supersymmetric descendant---in other words, a D-term---of a multiplet of composite operators free of contact
terms.
Our results were based on theories whose supercurrent multiplet satisfies the 
$(1,0)$ and $(1,1)$ conservation equations
\eqref{supercurrents-10} and \eqref{supercurrents-11}, respectively.
It would be very interesting to extend our analysis to the case of $(p,q)$ extended supersymmetry. 
For example, an analysis to appear \cite{Chang:2019kiu,SUSYDBITTbar}
of the $\cN=(2,2)$ case indicates that the strategy we have adopted of writing 
$O$ as a D-term fails for the most general $\cN=(2,2)$ supercurrent
which is described by the $\cS$-multiplet studied in \cite{Dumitrescu:2011iu}. 
Moreover, in the case of more supersymmetry, from a dimensional argument it seems likely that $O$ should be the bottom component of a suitable short supersymmetric multiplet.
Therefore, a more detailed analysis of the structure of supercurrents and short multiplets is in order.
It is interesting to point out that the so-called $T\bar{J}$ and $J\bar{T}$ deformations
\cite{Guica:2017lia} also preserve supersymmetry, at least in the $(1,0)$, $(2,0)$ and $(1,1)$ cases
\cite{SUSYJTbar}.
For these deformations, and also for subclasses of $(2,2)$ $T\bar{T}$ deformations, 
 the primary operator is not of Smirnov-Zamolodchikov type, see eq.~\eqref{AABB}.
Nonetheless, supersymmetric extensions of the arguments used in \cite{Smirnov:2016lqw} show that the resulting 
composite operators are free of short-distance singularities as well,
 See \cite{Jiang:2019hux} for the first example of this phenomena in the $\cN=(2,0)$ case.
We will report soon on these topics, which 
are currently under investigation, in future publications \cite{Jiang:2019hux,Chang:2019kiu,SUSYDBITTbar,SUSYJTbar}.

It is also rather striking that, when coupling the action to linearised supergravity, the $T\bar{T}$ deformation takes the form of a shift of the prepontential superfields. This fits well with the proposal that these deformations can be related to Jackiw-Teitelboim gravity~\cite{Dubovsky:2017cnj, Dubovsky:2018bmo} and to the geometry of the two-dimensional space~\cite{Conti:2018tca}, and indeed suggests that it might be possible to extend such relations to superspace.

As a simple but non-trivial example, we have constructed the $T\bar{T}$ deformation of a free theory of eight 
$\mathcal{N}=(1,1)$ scalar multiplets. 
We have done so by exploiting a map between $T\bar{T}$ deformations and (super)strings in light-cone gauge first highlighted in ref.~\cite{Baggio:2018gct}. We constructed the deformation of a \textit{free}, relativistic theory from superstrings in \textit{flat} space. It is natural to ask which theories might correspond to non-flat (super)string backgrounds. A first step in this direction was taken in ref.~\cite{Baggio:2018gct}, where it was argued that AdS$_3$ backgrounds supported by Neveu-Schwarz-Neveu-Schwarz fluxes \textit{only} are closely related to the deformation of  a  \textit{free non-relativistic} theory, see also ref.~\cite{Dei:2018mfl}. The appearance of a non-relativistic two-dimensional dynamics is rather common in uniform light-cone gauge, see \textit{e.g.}\ the review~\cite{Arutyunov:2009ga}. This should not pose an obstacle to studying $T\bar{T}$ deformations, as these have been recently generalised to non-relativistic theories~\cite{Cardy:2018jho}. Hence, it would  be very interesting to investigate this relation with string theory in greater detail.
Another direction which deserves further attention is how our uniform light-cone gauge construction of $T\bar{T}$ deformations might relate to the construction of such deformations in terms of gravity~\cite{Dubovsky:2017cnj, Dubovsky:2018bmo} and background geometry~\cite{Conti:2018tca}.
We hope to return to some of these questions in the future.

\section*{Acknowledgements}
We thank Andrea Cavagli\`a for collaboration at the initial stages of this project.
We also thank Nikolay Bobev, Sergei Dubovsky, Guzm\'an Hern\'andez-Chiffet, Sylvester James Gates Jr., 
Victor Gorbenko, Sergei Kuzenko, Edoardo Lauria, Mark Mezei, Stefano Negro, Ben Hoare and Stijn van Tongeren for useful related discussions. We are especially grateful to Andrea Cavagli\`a and Roberto Tateo for their comments on a preliminary version of this article. This work is partially supported through a research grant of the Swiss National Science Foundation, as well as by the NCCR SwissMAP, funded by the Swiss National Science Foundation.
The work of MB is supported by the European Union's Horizon 2020 research and innovation programme under the
Marie Sk{\l}odowska-Curie grant agreement no.~665501 with the Research Foundation Flanders (FWO). MB is an FWO [PEGASUS]${}^2$ Marie
Sk{\l}odowska-Curie Fellow.
The work of GT-M was supported by the
Interuniversity Attraction Poles Programme initiated by the Belgian Science Policy (P7/37),
 by the COST Action MP1210,
by the KU Leuven C1 grant ZKD1118 C16/16/005,
by the Albert Einstein Center for Fundamental Physics, University of Bern,
and by the Australian Research Council (ARC) Future Fellowship FT180100353.
GT-M thanks for support also the University of Western Australia, the University of Queensland and 
the University of Melbourne during the final stages of this work.
MB, AS and GT-M  thank the participants of the workshop \textit{A fresh look at $AdS_3/CFT_2$} in Villa Garbald, Castasegna, for the stimulating atmosphere where this work was started.

\appendix
\section{Conventions}
\label{app:conventions}
Below we describe our conventions.

\subsection{Two-dimensional conventions}
\label{app:conventions2d}
The two-dimensional metric is $\eta_{ab}=\text{diag}(-1,+1)$. The Levi-Civita tensor satisfies $\epsilon^{01}=+1$. We introduce light-cone coordinates $\sigma^{++}$ and $\sigma^{--}$ as
\begin{equation}
\label{eq:lc2d}
\sigma^{\pm\pm}= \frac{1}{2}\big(\sigma^0\pm\sigma^1\big)\,,
\end{equation}
so that for a co-vector $V_a$ we have
\begin{equation}
V_{\pm\pm}=\big( V_0\pm V_1\big)\,.
\end{equation}
The light-cone metric is then 
$\eta_{\pm\pm,\pm\pm}=\eta^{\pm\pm,\pm\pm}=0$,
$\eta_{\pm\pm,\mp\mp}=-2$,
$\eta^{\pm\pm,\mp\mp}=-1/2$.

\subsection{Ten-dimensional conventions}
\label{app:conventions10d}
The ten-dimensional metric is
\begin{equation}
\eta_{\mu\nu}= \text{diag}(-1,+1,\dots +1)\,.
\end{equation}
The ten-dimensional Gamma matrices  satisfy
\begin{equation}
(\Gamma^0)^{\text{t}}=-\Gamma^0,\qquad
(\Gamma^i)^{\text{t}}=\Gamma^i,\quad i=1,\dots 9,
\end{equation}
and can be written as
\begin{equation}
\Gamma^0 = i\sigma^{(2)}\otimes \mathbb{I}_{16},\qquad \Gamma^{i}=\sigma^{(1)}\otimes \gamma^i,\quad i=1,\dots9,\qquad
\Gamma_{11}=\sigma^{(3)}\otimes \mathbb{I}_{16}\,,
\end{equation}
where $\gamma^{i}$, $i=1,\dots8$ are 8-dimensional Euclidean Gamma matrices and $\gamma^9=\gamma^1\cdots\gamma^8$.
The charge conjugation is given by $\Gamma^0$, so that the Majorana condition forces the components of $\theta^I$ to be real, and $\bar{\theta}^{I}=(\theta^I)^{\text{t}}\Gamma^0$. For the Weyl condition, we impose $\tfrac{1}{2}(\mathbb{I}+\Gamma_{11})\theta=\theta$ which for type IIB strings can be solved by
\begin{equation}
\theta^1=\frac{1}{2}\left(\begin{array}{c}
\psi_-^i\\
0
\end{array}\right),\qquad
\theta^2=\frac{1}{2}\left(\begin{array}{c}
\psi_+^i\\
0
\end{array}\right),
\end{equation}
where $\psi_{\pm}^i$ have 16 real entries and the normalisation is chosen for future convenience.

\section{Derivation of deformed \texorpdfstring{$\mathcal{N}=(1,0)$}{N=(1,0)} actions}
\label{app:onecommazeroaction}
We collect here the derivation of some results for $\mathcal{N}=(1,0)$ actions.
\subsection{The Noether stress-energy tensor and deformation}
\label{app:nother10}
We start by recalling the definition of the Noether stress-energy tensor~$T_{ab}^{\text{(N)}}$
\begin{equation}
T_{ab}^{\text{(N)}}=\sum_i\eta_{ac}\frac{\delta \mathcal{L}}{\delta \partial_c\Phi_i}\partial_b\Phi_i - \eta_{ab}\mathcal{L}\,,
\end{equation}
where $\Phi_i$ are all the fields of the theory. In particular, for the free action~\eqref{eq:free10action} we have, in the light-cone coordinates~\eqref{eq:lc2d}
\begin{equation}
\begin{aligned}
&T_{++,++}^{\text{(N)}}=-\partial_{++}X\partial_{++}X -\ri\,\psi_{+}\partial_{++}\psi_+,\qquad
&&T_{++,--}^{\text{(N)}}=0,\\
&T_{--,++}^{\text{(N)}}=\ri\,\psi_{+}\partial_{--}\psi_+,\qquad
&&T_{--,--}^{\text{(N)}}=-\partial_{--}X\partial_{--}X,
\end{aligned}
\end{equation}
Note that this is not a symmetric traceless tensor unless we use the fermion equations of motion~$\partial_{--}\psi_+=0$.
It is convenient to rewrite the definition of the Noether stress-energy tensor in terms of a linear operator
\begin{equation}
\mathbf{K}_{ab}[\mathcal{X}]=\frac{1}{2}\sum_i\eta_{ac}\frac{\delta \mathcal{X}}{\delta \partial_c\Phi_i}\partial_b\Phi_i\,.
\end{equation}
Then, for the terms that make up the action~\eqref{eq:ansatz10} we find
\begin{equation}
\begin{aligned}
\mathbf{K}_{ab}[F(x)]=&-
\left(
\begin{array}{ccc}
\partial_{++}X\partial_{++}X &\ & \partial_{--}X\partial_{++}X\\
\partial_{--}X\partial_{++}X &\ &\partial_{--}X\partial_{--}X
\end{array}
\right)\,\alpha\,F'(\var)\,,\\
\mathbf{K}_{ab}[\psi_+\partial_{--}\psi_+]=&-
\left(
\begin{array}{ccc}
\psi_+\partial_{++}\psi_+ && \psi_+\partial_{--}\psi_+\\
0 &\ & 0
\end{array}
\right),\\
\mathbf{K}_{ab}[(\partial_{--}X)^2\psi_+\partial_{++}\psi_+]=&-
\left(
\begin{array}{ccc}
2\partial_{++}X\, \psi_+\partial_{++}\psi_+ &\ & 2\partial_{--}X\, \psi_+\partial_{++}\psi_+\\
\partial_{--}X\, \psi_+\partial_{++}\psi_+ && \partial_{--}X\, \psi_+\partial_{--}\psi_+
\end{array}
\right)\,\partial_{--}X\,.\\
\end{aligned}
\end{equation}
Using these expressions we can easily find the stress-energy tensor as a function of $A(\var)$, $B(\var)$, $C(\var)$ and their derivatives. We find that
\begin{equation}
\begin{aligned}
\text{det}[T^{\text{(N)}}]=&-\frac{1}{\alpha^2}A(A-2\var \,A')+\frac{1}{\alpha}\big[\var A'(\var C+B)-A(B-2\var B')\big]\,\ri\psi_+\partial_{--}\psi_+\\
&+\big[A'(B+\var C)+A(C+2\var C)\big]\,(\partial_{--}X)^2\, \ri\psi_+\partial_{++}\psi_+\,.
\end{aligned}
\end{equation}
Hence, from eq.~\eqref{eq:TTbardiffeq} we have
\begin{equation}
\begin{gathered}
0=A(1-A)+\var A'(2A-1),\qquad
0=\var A'(B+\var C)-\var B'-A(B-2\var B'),\\
0=A'B-C(1-A-\var A')-\var C'(1-2A),
\end{gathered}
\end{equation}
and one can verify that indeed~\eqref{eq:solNoether} gives a solution.

\subsection{Direct construction of the supersymmetric action}
It is quite easy, for this particular case, to construct an action that is invariant under eq.~\eqref{eq:susyvar}. In fact, following the logic of appendix~\ref{app:nother10}, we get
\begin{equation}
\begin{aligned}
\frac{\delta}{\delta\epsilon_-} F(\var)=&\,\ri\alpha(\partial_{++}X\partial_{--}\psi_+ + \partial_{++}\psi_+\partial_{--}X)\,F'(x),
\end{aligned}
\end{equation}
while the variations of the fermion bilinears follow immediately from~\eqref{eq:susyvar}. After taking the variation, we get two conditions which multiply the cubic and linear fermion terms; they are, respectively,
\begin{equation}
B'=2C+\var\,C'\,,\qquad
A'=B\,.
\end{equation}
This allows us to write down the supersymmetric version of \textit{any} bosonic Lagrangian given by~$A(\var)$. It is easy to verify that eq.~\eqref{eq:solSusy} solves these equations.

\subsection{Improved stress-energy tensor}
\label{app:improved}
We want here to construct a stress-energy tensor for the $\mathcal{N}=(1,0)$ theory which is symmetric and, for a conformal theory, traceless. We can do so by following the Belinfante procedure. Formally, we can treat $\psi_+$ as if it were a complex field, introducing fermion bilinears $\psi_+^* \partial_{\pm\pm} \psi_+$ and coupling the theory to a curved metric and spin-connection. Then the variations of the spin-connection contribution precisely yields the needed improvement terms. In practice, this boils down to modifying the linear operator~$\mathbf{K}$ of appendix~\ref{app:nother10} as
\begin{equation}
\begin{aligned}
\mathbf{K}_{ab}[\psi_+\partial_{--}\psi_+]=&-
\left(
\begin{array}{ccc}
\psi_+\partial_{++}\psi_+ && \psi_+\partial_{--}\psi_+\\
\psi_+\partial_{--}\psi_+ &\ & 0
\end{array}
\right),\\
\mathbf{K}_{ab}[\psi_+\partial_{++}\psi_+]=&-
\left(
\begin{array}{ccc}
0 &\ & \psi_+\partial_{++}\psi_+\\
\psi_+\partial_{++}\psi_+ &&  \psi_+\partial_{--}\psi_+
\end{array}
\right)\,.
\end{aligned}
\end{equation}
In this way we get to a set of differential equations for the coefficients $A(\var)$, $B(\var)$ and $C(\var)$, which read
\begin{equation}
\begin{gathered}
0=(1+A)A-\var A'(1+2A),\qquad
0=\var A'C+B'(1+2A),\\
0=A'B+(1+2A)C+\var C'(1+2A).
\end{gathered}
\end{equation}
The solution of this is given by eq.~\eqref{eq:improved}.

\section{Flat-space Green-Schwarz action in uniform light-cone gauge}
\label{app:GS}
We collect here some further details on the computation of the light-cone gauge-fixed Lagrangian for Green-Schwarz strings in flat space.

\subsection{Light-cone coordinates and Gamma matrices}
\label{app:lcgauge}
We start by introducing the projectors
\begin{equation}
G^\pm = \frac{1}{2}(\Gamma^9\pm \Gamma^0)
\,.
\end{equation}
 These have rank 16 and satisfy
\begin{equation}
G^\pm G^\pm = 0,\qquad
G^+G^- + G^+ G^-=1,\qquad
(G^\pm)^{\text{t}}=G^{\mp}\,.
\end{equation}
The $\kappa$-gauge fixing~\eqref{eq:kappagauge} is then $G^+\theta^I=0$. Notice that the projectors are related to the target-space Gamma matrices as
\begin{equation}
\Gamma^+=a\,\Gamma^9+(1-a)\Gamma^0=G^++(2a-1)G^-= G^++2\alpha\,G^-,\qquad
\Gamma^-=\Gamma^9-\Gamma^0=2\,G^-\,.
\end{equation}
The parameter~$\alpha$ appears naturally to measure the deviation of~$\Gamma^+$ from the light-cone. Similarly, the metric in the light-cone directions is given by
\begin{equation}
\eta_{\mu\nu}=\left(
\begin{array}{ccc}
g_{++}&g_{+-}&0\\
g_{+-}&g_{--}&0\\
0&0&\delta_{ij}
\end{array}\right)\,,
\end{equation}
with
\begin{equation}
g_{++}=0,\,\quad g_{+-}=1,\quad g_{--}=-2\alpha\,,\qquad
g^{++}=2\alpha,\quad g^{+-}=1,\quad g^{--}=0\,.
\end{equation}
\subsection{Fermion bilinears}
\label{app:bilinears}
The $\kappa$ gauge fixing imposes
\begin{equation}
G^+ \theta^I=(\theta^I)^{\text{t}}G^-=\bar\theta^I G^+ = 0\,,
\quad	 I=1,2\,.
\end{equation}
Let us observe that under this condition we can simplify a term of the form $\bar{\theta}^I \Gamma^\mu \theta^I$, which is what appears in the Green-Schwarz action~\eqref{eq:GSaction}. We have
\begin{equation}
\bar{\theta}^I \Gamma^\mu \theta^J=\begin{cases}
\ 0&\mu=1,\dots 8\,,\\
\ 2\,\phantom{\alpha}\,\big(\theta^I\big)^\text{t}\, \theta^J &\mu=-\,,\\
\ 2\,\alpha\,
\big(\theta^I\big)^\text{t}\, \theta^J\qquad\qquad&\mu=+\,.
\end{cases}
\end{equation}
It is therefore natural to introduce the short-hands~\eqref{eq:fermionshorthand}.

\subsection{T-duality and uniform light-cone gauge fixing}
\label{app:Tduality}
In order to impose the gauge-fixing condition $P_-=1$ from eq.~\eqref{eq:lcgauge} without having to perform a Legendre transform, it is convenient to T-dualise the coordinate~$X^-$, or in other words to gauge the $X^-$ isometry. This can be done for a generic background by modifying the string NLSM Lagrangian $\mathcal{L}$ by introducing the gauge field $A_a$:
\begin{equation}
\mathcal{L}(\partial_a X^+, \partial_a X^-, \partial_a X^i)
\to
\mathcal{L}_{\text{new}}=\mathcal{L}(\partial_a X^+, \partial_a X^-+A_a, \partial_a X^i) + \widetilde{X}^{-} \epsilon^{ab}\partial_b A_a\,.
\end{equation}
The equations of motion for~$\widetilde{X}^{-}$ impose that $\epsilon^{ab}\partial_a A_b=0$ \textit{i.e.}\ that the gauge connection is flat. In this way the action is unchanged, \textit{i.e.}\ $\int \mathcal{L}=\int\mathcal{L}_{\text{new}}$. On the other hand we have
\begin{equation}
\frac{\delta \mathcal{L}_{\text{new}}}{\delta{A_0}}=\frac{\delta \mathcal{L}_{\text{new}}}{\delta\partial_0 X^-}-\partial_1\widetilde{X}^{-}= P_- -\partial_1\widetilde{X}^{-}\,,
\end{equation}
so that the equations of motion for $A_0$ give $\partial_1 \widetilde{X}^{-}=P_-$ and we can implement the gauge fixing~\eqref{eq:lcgauge} by
\begin{equation}
\label{eq:tdualgauge}
X^+=\sigma^0,\qquad\widetilde{X}^{-}=\sigma^1\,.
\end{equation}
Working out the T-dual action and gauge-fixing is relatively straightforward for flat space, and it can be done \textit{e.g.}\ following ref.~\cite{Arutyunov:2014jfa}.

\bibliographystyle{JHEP}
\bibliography{refs}

\end{document}